\documentstyle[epsfig,ltwol]{article}

\def\be{\begin{equation}}
\def\ee{\end{equation}}
\def\bea{\begin{eqnarray}}
\def\eea{\end{eqnarray}}

\begin{document}

\title{
\hspace{4.5in}\parbox[t]{2in}{CMU-HEP98-07\\
DOE-ER/40682-103\\}
\\
\vspace*{.1in}
First Charm Hadroproduction Results from SELEX$^\dag $}

%
%
%
%
%
%

\author{\large The \textsc{Selex} Collaboration\\}
\author{\small\noindent
  J.~Russ$^{3}$,
  N.~Akchurin$^{17}$,
  V.~A.~Andreev$^{11}$,
  A.G.~Atamantchouk$^{11}$,
  M.~Aykac$^{17}$,
  M.Y.~Balatz$^{8}$,
  N.F.~Bondar$^{11}$,
  A.~Bravar$^{22}$,
  M.~Chensheng$^{7}$,
  P.S.~Cooper$^{5}$
% ...
\\
}

\address{\noindent\footnotesize
% ...
$^{3}$ Carnegie-Mellon University, Pittsburgh, PA 15213, U.S.A.\\
% ...
$^{5}$ Fermilab, Batavia, IL 60510, U.S.A.\\
% ...
$^{7}$ Institute of High Energy Physics, Beijing, PR China\\
$^{8}$ Institute of Theoretical and Experimental Physics, Moscow, Russia\\
% ...
$^{11}$ Petersburg Nuclear Physics Institute, St. Petersburg, Russia\\
% ...
$^{17}$ University of Iowa,  Iowa City, Iowa  52242, U.S.A.\\
% ...
$^{22}$ University of Trieste and INFN, Trieste, Italy\\
}


\twocolumn[\maketitle\abstracts{The SELEX experiment (E781) at Fermilab
is a 3-stage magnetic spectrometer 
for the high statistics study of charm hadroproduction out to
   large $x_F$ using 600 GeV $\Sigma^-$, p and $\pi$ beams. The main features 
of the spectrometer are: 
\begin{itemize}
\item high precision silicon vertex system
\item  broad-coverage particle identification with  TRD and RICH
\item  3-stage lead glass photon detector
\end{itemize}
   Preliminary results on differences in hadroproduction characteristics of 
charm mesons and $\Lambda_c^+$ for $x_F \ge 0.3$ are reported.  For baryon 
beams there is a striking asymmetry in the production of baryons 
compared to antibaryons.  
Leading particle effects for all incident hadrons are discussed.}]

\section{Introduction}
Understanding charm hadroproduction at 
fixed-target energies has been a difficult theoretical problem
because of the complexities of renormalization scale, of parton scale, 
and of hadronization 
corrections.  The recent review by Frixione, Mangano, Nason, and Ridolfi 
summarizes the theoretical
situation, using data through 1996~\cite{nason}.  More recent 
data from Fermilab E791 (500 GeV $\pi^-$
beam) greatly improves the statistical precision on charm meson production 
by pions, but E791 has not yet 
reported absolute cross sections or compared yields between charm species.
In this first report of the 
SELEX hadroproduction results, we compare our pion results at 580 GeV with 
those from E791 as well as comparing SELEX
pion data with our proton data at 550 GeV and $\Sigma^-$ data at 
620 GeV mean momenta.  All SELEX data
were taken in the same spectrometer with the same trigger.
We limit this report to data 
having $x_F \ge 0.3$, where the spectrometer acceptance is
essentially constant with $x_F$ for all final states.

\section{The Experiment}

\begin{figure}[htb]
 \begin{center}
\includegraphics[angle=270,width=3.4in, bb= 0 0 612 792]{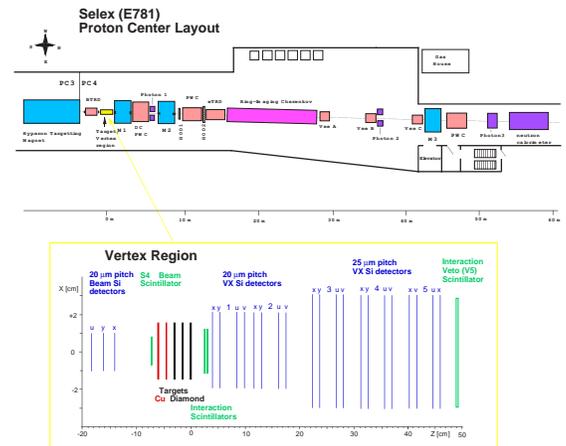}
\caption{E781 Layout}
\label{fig:spect}
\end{center}
\end{figure}

SELEX used the Fermilab Hyperon beam in negative polarity to make a mixed beam
 of $\Sigma$ and $\pi$ in
roughly equal numbers.  In positive polarity, protons comprised 92\% of the
 particles, with $\pi^+$
making up the balance.  The beam was run at 0 mrad production.  The experiment
 aimed especially at 
understanding charm production in the forward hemisphere and was built to have
 good mass and vertex 
resolution for charm momenta from 100-500 GeV/c.  The spectrometer is shown in 
Figure~\ref{fig:spect}.  

Interactions occurred in a target stack of 5 foils: 2 Cu and 3 C.  
Total target thickness was 
5\% of $\Lambda_{int}$ for protons.  Each foil was spaced by 1.5 cm from its 
neighbors.  Decays occurring
inside the volume of a target were rejected in this analysis.  Interactions 
were selected by a scintillator trigger.  The charm trigger was very loose,
 requiring only $\ge 4$ charged tracks in a forward $10^{\circ}$ cone and 
$\ge 2$ hits in a hodoscope after the second
analyzing magnet.  We triggered on about 1/3 of all inelastic interactions.

A major innovation in E781 was the use of online selection criteria to 
identify reconstructable events.  This experiment uses a RICH counter to 
identify $p$, $K$, or $\pi$ after the second analyzing magnet.  A computational
 filter used only these RICH-identifiable
tracks to make a full vertex reconstruction in the vertex silicon and
 downstream PWCs.  It selected events that had 
evidence for a secondary vertex.  This reduced the data size (and offline
 computation time) by a factor of nearly 8
at a cost of about a factor of 2 in charm written to tape, as normalized from
 a study of unfiltered $K_s^0$ and $\Lambda^0$ decays.  Most of the charm loss
 came from selection
cuts that are independent of charm species or kinematic variables.  No bias is
 expected from the filter.  Filter operation depends on stable track
 reconstruction and detector alignment.  These features were
monitored online and were extremely stable throughout the run.

\section{Charm Selection}

All data reported here result from a preliminary pass through the data, using
 a production code optimized for 
speed but not efficiency.  Final yields will be higher 
than these preliminary results.
However, our simulations indicate that the inefficiency does not affect the
 kinematic features of the
results for $x_F \ge 0.3$.  For all final states, the charm selection required
 that the primary vertex lie
within the target region and that the secondary vertex occur before the start
 of the VX silicon.  At our
high energy, this latter cut removed a number of $D^{\pm}$ events which can be
 recovered later.

In this analysis secondary vertices were reconstructed when the vertex 
$\chi^2$ for the ensemble of tracks was 
inconsistent with a single primary vertex.  All combinations of tracks were 
investigated, and every secondary vertex 
candidate was tested against a reconstruction table that listed acceptable
particle identification tags for a charm candidate, track selection criteria
 necessary (RICH identification
for a proton, for example), and any other selections, e.g., minimum
 significance cut for primary/secondary
vertex separation.  Selected events were written to output files and the
 essential reconstruction features for
each identified secondary vertex were saved in a PAW-like output structure for
 quick pass-II analysis.  All
data shown here come from analysis using this reduced output.

\subsection{System performance for charm}

\begin{figure}[htb]
\begin{center}
\includegraphics[width=3.4in, bb = 40 150 560 660]{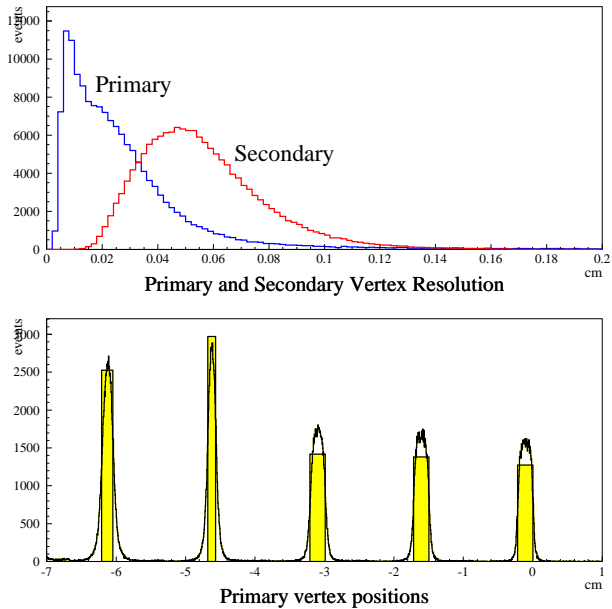}
\caption{Typical Primary and Secondary Vertex Error Distributions}
\label{fig:vtx}
\end{center}
\end{figure}

Vertex resolution is a critical factor in charm experiments.  The primary and
 secondary longitudinal vertex resolution for
all data in a typical run of the experiment are shown in Figure~\ref{fig:vtx}.
  The lower plot shows the
primary vertex distribution overlaid on rectangles that represent the physical
 placement of the 5 targets.
The average relativistic transformation factor from lab time to proper time for
 charm states in these data is
100.  This spatial resolution corresponds to about a 20 fs proper time
 resolution for lifetime studies.

\begin{figure}[htb]
\begin{center}
\includegraphics[width=3.4in, bb = 40 150 560 660]{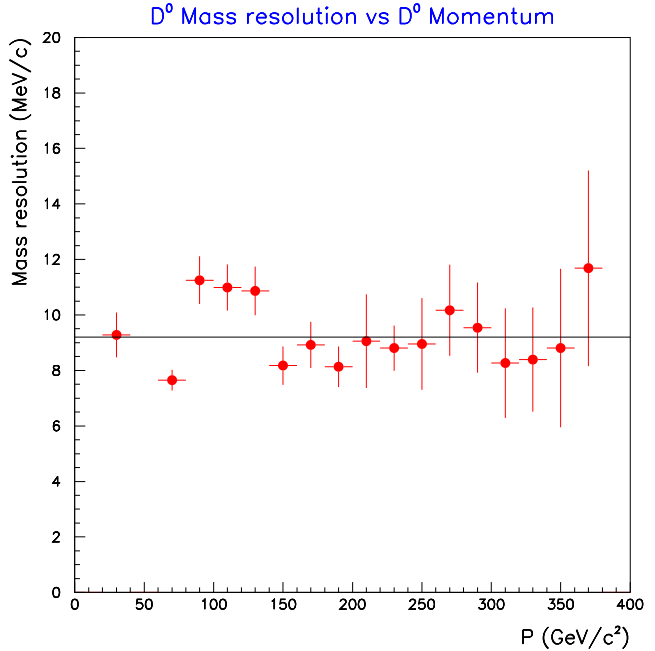}
\caption{$D^0$ Mass Resolution versus $D^0$ Momentum}
\label{fig:massres}
\end{center}
\end{figure}

Another important factor in charm studies at large $x_F$ is having good charm
 mass resolution at all momenta.
Figure~\ref{fig:massres} shows that the measured width of the $D^0 \rightarrow
 K^- + \pi^+$ is about 10 MeV
for all $x_F$.  

\begin{figure}[htb]
\epsfxsize=0.45\textwidth \epsfysize=0.3\textheight
\epsffile {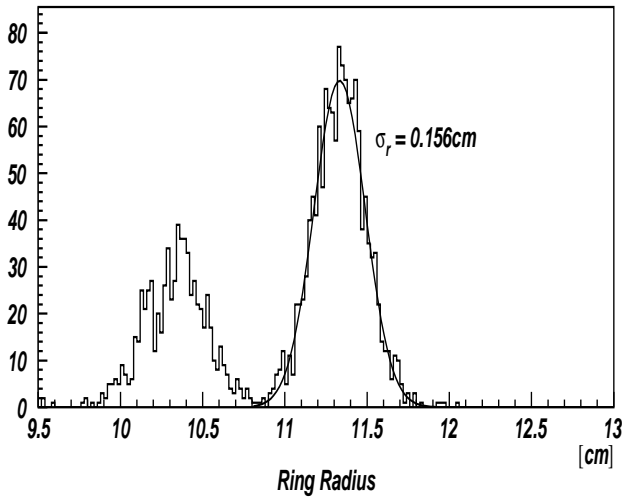}
\caption{RICH $K$ and $\pi$ Response at 100 GeV/c}
\label{fig:RICH}
\end{figure}

Finally, we depend on the RICH to give correct identification
 of K and p decay prongs.
Figure~\ref{fig:RICH} shows the $\pi$/K separation in interaction data for 
100 GeV/c tracks, a typical
momentum for prongs from our charm states.  The RICH gives 
$\pi$/K separation up to 165 GeV/c (2$\sigma$ confidence level)~\cite{richnim}.

\section{Overall Charm Features at Large $x_F$}

\begin{figure}
\begin{center}
\includegraphics[width=3.4in, bb = 10 150 560 690]{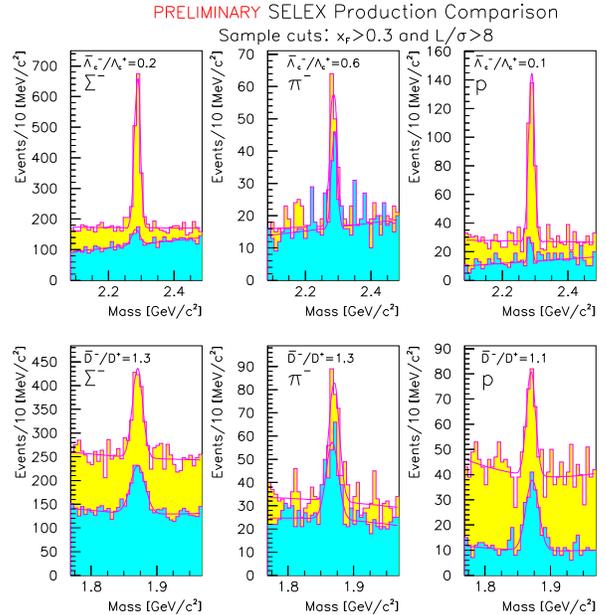}
\caption{Charm and Anticharm mass distributions for $\Sigma^-, \pi^-$,
and p beams in modes $\Lambda_c^+ \rightarrow p K^- \pi^+$ or c.c.\
and $D^+ \rightarrow K^- \pi^+ \pi^+$ or c.c.}
\label{fig:masses}
\end{center}
\end{figure}

Previous high-statistics charm production results from pions~\cite{pidata} and
 protons~\cite{protondata} have
emphasized central production, although both NA32 and E791 have presented 
results for $x_F \ge 0.5$.   SELEX and E769
are the only high energy experiments reporting results from three different
 beam particles with identical systematics.  
The important features of the SELEX data can be seen at a glance in 
Figure~\ref{fig:masses} for the charged
states $D^\pm$, $\Lambda_c^+$, and $\overline{\Lambda}_c^-$ produced 
respectively by $\Sigma^-$, $\pi^-$,
and proton beams.  The pion data show comparable particle and antiparticle 
yields both for charm mesons and for charm baryons, as reported by NA32 at 
lower energy~\cite{pidata}.  It remains a surprising feature of hadroproduction
that one finds significant antibaryon production from pions even at
$x_F \ge 0.5$.  The source of the antiquark pair which combines with
the charmed antiquark has been the subject of considerable theoretical
 speculation.  The pion provides a $\overline{u}$ valence quark which can
 contribute in some models.  No present model gives an adequate description.  
There is good agreement for the $D^{\pm}$ production asymmetry integrated over
 $x_F \ge 0.3$ between these preliminary results
and the E791 results~\cite{pidata}.
E791 has not published $\Lambda_c^+$ asymmetry 
results.  Their observations are consistent with these shown here~\cite{kwan}.

The relative efficiencies for each beam particle are almost the same in this
$x_F$ region, so
that one can quote the ratio of the cross sections even though we have not yet
determined absolute yields.  The normalization between
different incident hadrons depends on the number of incident beam particles 
for each data sample and on the total inelastic cross section for each
beam particle.  We use 34 mb for the proton inelastic cross section,
27 mb for $\Sigma^-$, and 22 mb for $\pi^-$ to compare yields for different 
beam particles.  For these data the relative yields of selected charmed
states, normalized to pion production, are given in Table~\ref{Tab:ratio}.
No errors are included in this preliminary analysis.  Note that this
table does not directly provide information about the relative yields
for the different charmed states.

\begin{table}[htb]
\begin{tabular}{ c c c c } \hline \hline
Relative Charmed & p & $\pi^-$ & $\Sigma^-$ \\
Particle Yields        &  & & \\ \hline
$\overline{\Lambda}_c^-$ & 0.25 & 1.0 & 1.1 \\
$\Lambda_c^+$ & 0.9  & 1.0 & 1.2  \\
$D^-$ & 0.4 & 1.0 & 0.8 \\
$D^+$ & 0.2 & 1.0 & 0.4 \\
\hline\hline
\end{tabular}
\caption{ Relative Charmed Particle Yields for $x_F \ge $ 0.3 versus beam type}
\label{Tab:ratio}
\end{table}

Perhaps the most surprising result from this table is the observation that
baryon beams are  very effective charm baryon producers, at least at large 
$x_F$.   Also, for the states listed here, the $\Sigma^-$ beam has yields 
comparable to pions, except for the non-leading case of the $D^+$.  We have
not yet compiled the yields for the c-s-q baryons, where we expect the
 $\Sigma^-$ beam relative yields will large.

The previous table gave the relative efficacy of each beam particle for 
producing a given charm state at large $x_F$.  It does \emph{not} compare 
relative yields of the different charm states for the same beam.  
As can be seen from
Figure~\ref{fig:masses}, there are strong asymmetries.  These are tabulated
in Table~\ref{Tab:asym}.  Again, errors are omitted at this stage of analysis.

\begin{table}[htb]
\begin{tabular}{ c c c c } \hline \hline
Yield Ratio & p & $\pi^-$ & $\Sigma^-$ \\ \hline
$\overline{\Lambda}_c^-$/$\Lambda_c^+$ & 0.1  & 0.6 & 0.2  \\
$D^-$/$D^+$ & 1.1 & 1.2 & 1.3 \\
\hline\hline
\end{tabular}
\caption{  Charmed Particle Antiparticle Ratios for $x_F \ge 0.3$ 
versus beam type}
\label{Tab:asym}
\end{table}

Table~\ref{Tab:asym}  shows for both baryon beams there are striking 
differences in production asymmetries for charm baryons
compared to the pion beam.  For charm mesons, that is not the case.  Baryon 
beams, which have no valence antiquarks, show strong suppression  of 
antibaryon production, compared to pions.  This 
feature was not observed by NA27 in 400 GeV pp collisions.  They
reported comparable baryon/antibaryon production but had only a few events,
 all in the central region.  No other proton data exist for charm baryons.  
The WA89 results for charm baryon production by $\Sigma^-$ are 
consistent with our findings~\cite{WA89}.

The $D^-$ and $\Lambda_c^+$ are leading hadrons in the sense that all 3 beam 
hadrons \emph{may} contribute at least one valence quark to the final state.
The large difference in the  $\Lambda_c^+$ asymmetry between the meson beam 
(largely symmetric) and the baryon beams (very asymmetric)
 is a new issue for charm hadroproduction analysis, which has assumed that 
there is a
universal baryon/meson fraction for all incident hadrons~\cite{nason}.

\section{Summary}

The SELEX experiment complements previous charm hadroproduction 
experiments by exploring different regions of
production phase space and by using different beams.
The early results already show some noteworthy new features
of charm production.  
Further studies of different states and details 
of single- and double-differential charm
production distributions are underway and will be reported at
meetings in the fall.  

Further analysis will extend the $x_F$ coverage down to about 0.1, 
to enhance overlap with other experiments
and to increase statistics.
Also, other charm baryon states are being analyzed and results will
be reported later.

\section*{Acknowledgements}
We are indebted to the technical staff at Fermilab and our home
institutions, especially B.~C.~LaVoy, D.~Northacker, F.~Pearsall, and J.~Zimmer
, for invaluable technical support.
This project was supported in part by
Bundesministerium f\"ur Bildung, Wissenschaft, Forschung und Technologie,
Consejo Nacional de Ciencia y Tecnolog\'{\i}a {\nobreak (CONACyT)},
Conselho Nacional de Desenvolvimento Cient\'{\i}fico e Tecnol\'ogico,
Fondo de Apoyo a la Investigaci\'on (UASLP),
Funda\c{c}\~ao de Amparo \`a Pesquisa do Estado de S\~ao Paulo (FAPESP),
the Israel Science Foundation founded by the Israel Academy of
  Sciences and Humanities,
Istituto Nazionale de Fisica Nucleare (INFN),
the International Science Foundation (ISF),
the National Science Foundation,NATO,
the Russian Academy of Science,
the Russian Ministry of Science and Technology,
the Turkish Scientific and Technological Research Board (T\"{U}B\.ITAK),
the U.S. Department of Energy, and
the U.S.-Israel Binational Science Foundation (BSF).

\footnotetext[0]{\noindent $^\dag$Invited talk presented at the 1998 
International Conference on High Energy Physics, 
Vancouver, B.C., Canada, July, 1998.}
\addtocounter{footnote}{-12}
\addtocounter{footnote}{1}
\footnotetext{Present address: Instituto
    de Fisica da Universidade Estadual de Campinas, UNICAMP, SP, Brazil.}
\addtocounter{footnote}{1}
\footnotetext{Now at Imperial College, London SW7 2BZ, U.K.}
\addtocounter{footnote}{1}
\footnotetext{deceased}
\addtocounter{footnote}{1}
\footnotetext{Present address: Dept. of Physics,
      Wayne State University, Detroit, MI 48201, U.S.A.}
\addtocounter{footnote}{1}
\footnotetext{Present address: Universit\"at Freiburg, 79104 Freiburg, Germany}
\addtocounter{footnote}{1}
\footnotetext{Present address:
      Physik-Department, Technische Universit\"at M\"unchen,
      85748 Garching, Germany}
\addtocounter{footnote}{1}
\footnotetext{Current Address:
      Instituto de Fisica Teorica da Universidade Estadual Paulista,
      S\~ao Paulo, Brazil}
\addtocounter{footnote}{1}
\footnotetext{Present address: Lucent Technologies,Naperville, IL}
\addtocounter{footnote}{1}
\footnotetext{Now at Max-Planck-Institut f\"ur Physik, M\"unchen, Germany}
\addtocounter{footnote}{1}
\footnotetext{Present address: Motorola Inc., Schaumburg, IL}
\addtocounter{footnote}{1}
\footnotetext{Generous support of Carnegie-Mellon University 
                                         is gratefully acknowledged.}
\addtocounter{footnote}{1}
\footnotetext{Present address: Deutsche Bank AG, 65760 Eschborn, Germany}

\end{document}